\documentclass{amsart}

\newtheorem{theorem}{Theorem}[section]

\theoremstyle{definition}
\newtheorem{definition}[theorem]{Definition}

\theoremstyle{remark}

\numberwithin{equation}{section}

\begin{document}
\title{ VECTOR COHERENT STATES WITH AN UNBOUNDED INVERSE FRAME OPERATOR}
\author{K.Thirulogasanthar}
\address{Department of Mathematics and Statistics, Concordia University,  
   7141 Sherbrooke Street West, Montreal, Quebec H4B 1R6, Canada }
\email{t$\_$kengat@alcor.concordia.ca}
\date{\today}
\begin{abstract} 
We present a class of vector coherent states in the domain $D\times D\times ....\times D$ (n-copies), where $D$ is the complex unit disc, using a specific class of  hermitian matrices. Further, as an example, we build vector coherent states in the unit disc by considering the unit disc as the homogeneous space of the group $SU(1,1)$.
\end{abstract}
\maketitle
\section{Introduction}
As it is well-known, coherent states, CS has applications in several branches of quantum physics (for example, see \cite{key19}). CS can be defined in several ways \cite{key2}. In this article we take the following definition.
\begin{definition}
Let $\mathfrak H$ be a Hilbert space with an orthonormal basis $\{\phi_{m}\}_{m=0}^{\infty}$ and $\mathbb C$ be the complex plane. For $z\in\mathfrak D$, an open subset of $\mathbb C$, the states
\begin{equation}
\mid z\rangle=\mathcal N(|z|)\sum_{m=0}^{\infty}\frac{z^{m}}{\sqrt{\rho(m)}}\phi_{m}
\label{1}\end{equation}
are said to form a set of CS if
\begin{enumerate}
\item[(a)]
The states $\mid z\rangle$ are normalized,
\item[(b)]
The states $\mid z\rangle$ give a resolution of the identity, that is
\begin{equation}
\int_{\mathfrak D}\mid z\rangle W(|z|)\langle z\mid d\mu=I\label{2}
\end{equation}
\end{enumerate}
where $\mathcal N(|z|)$ is the normalization factor, $\{\rho(m)\}_{m=0}^{\infty}$ is a sequence of nonzero positive real numbers,  $W(|z|)$ is a positive function called a weight function, $d\mu$ is an appropriately chosen measure and $I$ is the identity operator on $\mathfrak H$.\label{d1}
\end{definition}
In the literature, several interesting classes of CS were formed by changing the factors and parameters of (\ref{1})(for example, the deformed CS). In \cite{key31} we have extended the above definition to a class of vector coherent states, VCS by replacing the complex number $z$ by a matrix $Z$ and  gave several examples using quaternion matrices. In this article we develop another class of VCS with a particular type of hermitian matrix.
\section{Vector Coherent States with an Unbounded frame operator}
In this section we present VCS in the domain $D\times D\times ....\times D$ (n-copies), where $D=\{ z\in\mathbb C \mid |z|<1 \}$ is the complex unit disc, using a hermitian matrix.
\subsection{Vector coherent states}
 For $z_{1},z_{2},...,z_{n-1}\in D$, let $Z= ( z_{1},z_{2},...,z_{n-1} )$. Define the $n\times n$ hermitian matrix, $\mathcal Z$ as
\begin{equation}
\mathcal Z=\mathbb I_{n}+\left(\begin{array}{cc}
0&Z\\
\overline{Z}^{T}& 0\end{array}\right)_{n\times n}\label{un1}
\end{equation}
where $\overline{Z}^{T}$ is the complex conjugate transpose of $Z$. Let $\{\chi^{1},\chi^{2},...,\chi^{n}\}$ be the natural orthonormal basis of $\mathbb C^{n}$.Form the Hilbert space $\widetilde{\mathfrak H}=\mathbb C^{n}\otimes\mathfrak H$, where $\otimes$ is the tensor product.
Let $\psi_{m}=\frac{\phi_{m}}{m+1}$. The set, $\{\Phi_{qm}=\chi^{q}\otimes \psi_{m}\mid q=1...n,m=0...\infty\}$, is a basis of $\widetilde{\mathfrak H}$.
With the above set up we form the set of VCS in $L^{2}(D\times D\times...\times D,d\mu)$, where we take 
\begin{equation}
d\mu=r_{1}r_{2}...r_{n-1}dr_{1}d\theta_{1}dr_{2}d\theta_{2}...dr_{n-1}d\theta_{n-1}\label{un2}
\end{equation}
 by taking $z_{j}=r_{j}e^{i\theta_{j}},j=1,2,..n-1$,as 
\begin{equation}
|\mathcal Z,q\rangle=\mathcal N(|\mathcal Z |)\sum_{m=0}^{\infty}R(m)\mathcal Z^{m}\Phi_{qm},\hspace{1cm}q=1,2,...,n\label{un3}
\end{equation}
The number $\mathcal N=N(|\mathcal Z |)$ and the $n\times n$ matrix $R(m)$ have to be chosen suitably. First of all let us calculate $\mathcal Z^{m}$ by diagonalizing the matrix $\mathcal Z$. The eigenvalues of $\mathcal Z$ are $1,1+a$ and $1-a$ with multiplicities $n-2, 1$ and $1$ respectively, where $a=\|Z\|=\sqrt{r_{1}^{2}+r_{2}^{2}+...+r_{n-1}^{2}}$. A set of orthogonal normalized eigenvectors corresponding to the eigenvalue, 1 are
\begin{eqnarray*}
V_{1}&=&\frac{1}{\sqrt{b_{1}}}\left(\begin{array}{cccccccccc}0&-z_{2}&z_{1}&0&0&0&0&0&...&0\end{array}\right)\\
V_{2}&=&\frac{\sqrt{b_{1}}}{r_{1}\sqrt{b_{2}}}\left(\begin{array}{cccccccccc}0&-\frac{r_{1}^{2}}{b_{1}}z_{3}&-\frac{1}{b_{1}}z_{1}\overline{z_{2}}z_{3}&z_{1}&0&0&0&0&...&0\end{array}\right)\\
V_{3}&=&\frac{\sqrt{b_{2}}}{r_{1}\sqrt{b_{3}}}\left(\begin{array}{cccccccccc}0&-\frac{r_{1}^{2}}{b_{2}}z_{4}&-\frac{1}{b_{2}}z_{1}\overline{z_{2}}z_{4}&-\frac{1}{b_{2}}z_{1}\overline{z_{3}}z_{4}&z_{1}&0&0&0&...&0\end{array}\right)\\
V_{4}&=&\frac{\sqrt{b_{3}}}{r_{1}\sqrt{b_{4}}}\left(\begin{array}{cccccccccc}0&\frac{-r_{1}^{2}z_{5}}{b_{3}}&\frac{-z_{1}\overline{z_{2}}z_{5}}{b_{3}}&\frac{-z_{1}\overline{z_{3}}z_{5}}{b_{3}}&\frac{-z_{1}\overline{z_{4}}z_{5}}{b_{3}}&z_{1}&0&0&...&0\end{array}\right)\\
V_{5}&=&\frac{\sqrt{b_{4}}}{r_{1}\sqrt{b_{5}}}\left(\begin{array}{cccccccccc}0&\frac{-r_{1}^{2}z_{6}}{b_{4}}&\frac{-z_{1}\overline{z_{2}}z_{6}}{b_{4}}&\frac{-z_{1}\overline{z_{3}}z_{6}}{b_{4}}&\frac{-z_{1}\overline{z_{4}}z_{6}}{b_{4}}&\frac{-z_{1}\overline{z_{5}}z_{6}}{b_{4}}&z_{1}&0&...&0\end{array}\right)\\
&&...................................................................\\
V_{n-2}&=&\frac{\sqrt{b_{n-4}}}{r_{1}\sqrt{b_{n-3}}}\left(\begin{array}{ccccccccc}0&\frac{-r_{1}^{2}z_{n-1}}{b_{n-3}}&\frac{-z_{1}\overline{z_{2}}z_{n-1}}{b_{n-3}}&\frac{-z_{1}\overline{z_{3}}z_{n-1}}{b_{n-3}}&.&.&.&\frac{-z_{1}\overline{z_{n-2}}z_{n-1}}{b_{n-3}}&z_{1}\end{array}\right)\\
\end{eqnarray*}

where $b_{j}=r_{1}^{2}+r_{2}^{2}+....r_{j+1}^{2},\hspace{.1cm}j=1,2...n-3$. Eigenvectors corresponding to the eigenvalues $1+a$ and $1-a$ are
\begin{eqnarray*}
V_{n-1}&=&\frac{1}{\sqrt{2}a}\left(\begin{array}{ccccc}-a&\overline{z_{1}}&\overline{z_{2}}&...&\overline{z_{n-1}}\end{array}\right)\\
V_{n}&=&\frac{1}{\sqrt{2}a}\left(\begin{array}{ccccc}a&\overline{z_{1}}&\overline{z_{2}}&...&\overline{z_{n-1}}\end{array}\right)\\
\end{eqnarray*}
respectively. The set $\{V_{1},V_{2},...,V_{n}\}$ is an orthonormal set with unit vectors. Form a matrix $P$ by placing $V_{j}s$ as columns, i.e,
$$P=\left(\begin{array}{cccc}V_{1}^{T}&V_{2}^{T}&...&V_{n}^{T}\end{array}\right).$$
Then $P^{T}\mathcal Z  P=D$, a diagonal matrix with the eigenvalues over the diagonal. Thus we obtain $\mathcal Z^{m}=PD^{m}P^{\dagger}=\left(c_{lj}\right)_{n\times n}$, where
\begin{eqnarray*}
c_{11}&=&E_{m}\\
c_{l1}&=&\frac{O_{m}\overline{z_{l-1}}}{a},\hspace{.2cm}l=2,3,...,n\\
c_{1j}&=&\frac{O_{m}z_{j-1}}{a},\hspace{.2cm}j=2,3,...,n\\
c_{ll}&=&\frac{r_{l-1}^{2}E_{m}+a^{2}-r_{l-1}^{2}}{a^{2}},\hspace{.2cm}l=2,3,..,n\\
c_{lj}&=&\frac{z_{j-1}\overline{z_{l-1}}}{a^{2}}(E_{m}-1),\hspace{.2cm}l\not=j,l\not=1,j\not=1
\end{eqnarray*}
with
\begin{eqnarray}
E_{m}&=&\frac{1}{2}[(1+a)^{m}+(1-a)^{m}]\label{un4}\\
O_{m}&=&\frac{1}{2}[(1+a)^{m}-(1-a)^{m}].\label{un5}
\end{eqnarray}
Further notice that $(\mathcal Z^{m})^{\dagger}=\mathcal Z^{m}$. Before we choose $\mathcal N$ and $R(m)$ suitably, let us write $R(m)$ tentatively as
$R(m)=\left(\alpha_{lj}\right)$ where
$$\alpha_{11}=R_{m}e^{im\theta},\hspace{.2cm}\alpha_{ll}=S_{m}e^{im\theta}, l\not=1,\hspace{.2cm}\alpha_{lj}=0,l\not=j$$
with
$$\theta=\theta_{1}+\theta_{2}+...+\theta_{n-1}.$$
Consider
\begin{eqnarray*}
|\mathcal Z,1\rangle\langle\mathcal Z,1|&=& \mathcal N^{2}\mid\sum_{m=0}^{\infty}R(m)\mathcal Z^{m}\Phi_{1m}\rangle\langle\sum_{k=0}^{\infty}R(k)\mathcal Z^{k}\Phi_{1k}\mid\\
&=&\mathcal N^{2}\sum_{m=0}^{\infty}\sum_{k=0}^{\infty}\mid R(m)\mathcal Z^{m}\Phi_{1m}\rangle\langle R(k)\mathcal Z^{k}\Phi_{1k}\mid\\
&=&\mathcal N^{2}\sum_{m=0}^{\infty}\sum_{k=0}^{\infty}R(m)\mathcal Z^{m}\Omega_{1}\left(\mathcal Z^{k}\right)^{\dagger}\left(R(k)\right)^{\dagger}\otimes\mid\psi_{m}\rangle\langle\psi_{k}\mid
\end{eqnarray*}
where $\Omega_{1}=\left(s_{lj}^{(1)}\right)_{n\times n}$ with $s_{11}^{(1)}=1$ and $s_{lj}^{(1)}=0$ for $l\not=1,j\not=1$. The matrix multiplication yields,
$$R(m)\mathcal Z^{m}\Omega_{1}\left(\mathcal Z^{k}\right)^{\dagger}\left(R(k)\right)^{\dagger}=\left(f_{lj}^{(1)}\right)_{n\times n}$$
where
\begin{eqnarray*}
f_{11}^{(1)}&=&R_{m}R_{k}E_{m}E_{k}e^{i\theta(m-k)}\\
f_{ll}^{(1)}&=&S_{m}S_{k}O_{m}O_{k}\frac{r_{l-1}^{2}}{a^{2}}e^{i\theta(m-k)},\hspace{.5cm}l=2,3,...,n\\
f_{1l}^{(1)}&=&R_{k}E_{k}S_{m}O_{m}\frac{r_{l-1}}{a}e^{i(m\theta-k\theta+\theta_{l-1})},\hspace{.5cm}l=2,3,...,n\\
f_{l1}^{(1)}&=&R_{k}E_{k}S_{m}O_{m}\frac{r_{l-1}}{a}e^{i(m\theta-k\theta-\theta_{l-1})},\hspace{.5cm}l=2,3,...,n\\
f_{lj}^{(1)}&=&S_{m}S_{k}O_{m}O_{k}r_{j-1}r_{l-1}e^{i(m\theta-k\theta+\theta_{j-1}-\theta_{l-1})},\hspace{.5cm}j,l=2,3,...,n;j\not=l.\\
\end{eqnarray*}
Thus we obtain
\begin{equation}
|\mathcal Z,1\rangle\langle\mathcal Z,1|=\mathcal N^{2}\sum_{m=0}^{\infty}\sum_{k=0}^{\infty}\left(f_{lj}^{(1)}\right)_{n\times n}\otimes\mid\psi_{m}\rangle\langle\psi_{k}\mid.\label{un6}
\end{equation}
Similarly, in order to obtain $|\mathcal Z,2\rangle\langle\mathcal Z,2|$ we calculate
$$R(m)\mathcal Z^{m}\Omega_{2}\left(\mathcal Z^{k}\right)^{\dagger}\left(R(k)\right)^{\dagger}=\left(f_{lj}^{(2)}\right)_{n\times n}$$
where
\begin{eqnarray*}
f_{11}^{(2)}&=&R_{m}R_{k}O_{m}O_{k}\frac{r_{1}^{2}}{a^{2}}e^{i\theta(m-k)}\\
f_{22}^{(2)}&=&S_{m}S_{k}\frac{(r_{1}^{2}E_{m}+a^{2}-r_{1}^{2})}{a^{4}}(r_{1}^{2}E_{k}+a^{2}-r_{1}^{2})e^{i\theta(m-k)}\\
f_{ll}^{(2)}&=&S_{k}S_{m}\frac{(E_{m}-1)(E_{k}-1)}{a^{4}}r_{1}^{2}r_{l-1}^{2}e^{i\theta(m-k)},\hspace{.5cm}l=3,4,...,n\\
f_{lj}^{(2)}&=&A_{ljmk}^{(2)}e^{i(m\theta-k\theta-\theta_{l-1})},\hspace{.5cm}j=1,l=2,3,...,n\\
f_{lj}^{(2)}&=&B_{ljmk}^{(2)}e^{i(m\theta-k\theta+\theta_{j-1})},\hspace{.5cm}l=1,j=2,3,...,n;j\not=l.\\
f_{lj}^{(2)}&=&C_{ljmk}^{(2)}e^{i(m\theta-k\theta+\theta_{j-1}-\theta_{l-1})},\hspace{.5cm}{\text{otherwise}}
\end{eqnarray*}
Where $A_{ljmk}^{(2)},B_{ljmk}^{(2)}$ and $C_{ljmk}^{(2)}$ are some functions of $m,k$ and $r_{l},r_{j}$s, exact expressions are irrelevant for our purpose.Thus we obtain
\begin{equation}
|\mathcal Z,2\rangle\langle\mathcal Z,2|=\mathcal N^{2}\sum_{m=0}^{\infty}\sum_{k=0}^{\infty}\left(f_{lj}^{(2)}\right)_{n\times n}\otimes\mid\psi_{m}\rangle\langle\psi_{k}\mid.\label{un7}
\end{equation}
In general, in order to get $|\mathcal Z,q\rangle\langle\mathcal Z,q|$, we calculate
$$R(m)\mathcal Z^{m}\Omega_{q}\left(\mathcal Z^{k}\right)^{\dagger}\left(R(k)\right)^{\dagger}=\left(f_{lj}^{(q)}\right)_{n\times n}$$
where
\begin{eqnarray*}
f_{11}^{(q)}&=&R_{m}R_{k}O_{m}O_{k}\frac{r_{q-1}^{2}}{a^{2}}e^{i\theta(m-k)}\\
f_{22}^{(q)}&=&S_{m}S_{k}\frac{(r_{q-1}^{2}E_{m}+a^{2}-r_{q-1}^{2})}{a^{4}}(r_{q-1}^{2}E_{k}+a^{2}-r_{q-1}^{2})e^{i\theta(m-k)}\\
f_{ll}^{(q)}&=&S_{k}S_{m}\frac{(E_{m}-1)(E_{k}-1)}{a^{4}}r_{q-1}^{2}r_{l-1}^{2}e^{i\theta(m-k)},\hspace{.5cm}l\not=q,l=2,4,...,n\\
f_{lj}^{(q)}&=&A_{ljmk}^{(q)}e^{i(m\theta-k\theta-\theta_{l-1})},\hspace{.5cm}j=1,l=2,3,...,n\\
f_{lj}^{(q)}&=&B_{ljmk}^{(q)}e^{i(m\theta-k\theta+\theta_{j-1})},\hspace{.5cm}l=1,j=2,3,...,n;j\not=l.\\
f_{lj}^{(q)}&=&C_{ljmk}^{(q)}e^{i(m\theta-k\theta+\theta_{j-1}-\theta_{l-1})},\hspace{.5cm}{\text{otherwise}}
\end{eqnarray*}
Thus we obtain
\begin{equation}
|\mathcal Z,q\rangle\langle\mathcal Z,q|=\mathcal N^{2}\sum_{m=0}^{\infty}\sum_{k=0}^{\infty}\left(f_{lj}^{(q)}\right)_{n\times n}\otimes\mid\psi_{m}\rangle\langle\psi_{k}\mid.\label{un8}
\end{equation}
Now, when we compute
$$\sum_{q=1}^{n}\int_{D\times...\times D}\mid\mathcal Z,q\rangle\langle\mathcal Z,q\mid d\mu,$$
all the off-diagonal terms vanish with one of the $\theta_{1},\theta_{2},...,\theta_{n-1}$ integrals and only in the diagonals terms with $m=k$ survive. Thus we obtain
$$\sum_{q=1}^{n}\int_{D\times...\times D}\mid\mathcal Z,q\rangle\langle\mathcal Z,q\mid d\mu=(D_{lj})_{n\times n}$$
where
\begin{eqnarray*}
D_{11}&=&\sum_{q=1}^{n}f_{11}^{q}=\sum_{m=0}^{\infty}\left(\int_{D\times...\times D}\mathcal N^{2}R_{m}^{2}(E_{m}^{2}+O_{m}^{2})d\mu\right)\otimes\mid\psi_{m}\rangle\langle\psi_{m}\mid\\
D_{ll}&=&\sum_{q=1}^{n}f_{ll}^{q}\\
&=&\sum_{m=0}^{\infty}\left(\int_{D\times...\times D}\mathcal N^{2}S_{m}^{2}\frac{r_{l-1}^{2}(E_{m}^{2}+O_{m}^{2})+a^{2}-r_{l-1}^{2}}{a^{2}}d\mu\right) \otimes\mid\psi_{m}\rangle\langle\psi_{m}\mid,\hspace{.2cm}l=2,...,n\\
D_{lj}&=&0\hspace{.5cm}{\text{otherwise}}.
\end{eqnarray*}
Choose
\begin{equation}
\mathcal N=\frac{a}{\sqrt{(2\pi)^{n-1}}}.\label{un9}
\end{equation}
Then
$$D_{11}=\sum_{m=0}^{\infty}\left(\int_{0}^{1}...\int_{0}^{1}a^{2}R_{m}^{2}(E_{m}^{2}+O_{m}^{2})dr_{1}...dr_{n-1}\right)\otimes\mid\psi_{m}\rangle\langle\psi_{m}\mid$$
and
$$D_{ll}=\sum_{m=0}^{\infty}\left(\int_{0}^{1}...\int_{0}^{1}S_{m}^{2}(r_{l-1}^{2}(E_{m}^{2}+O_{m}^{2})+a^{2}-r_{l-1}^{2})dr_{1}...dr_{n-1}\right)\otimes\mid\psi_{m}\rangle\langle\psi_{m}\mid,\hspace{.2cm}l=2,3,...n.$$
The following integrals are finite and positive
\begin{eqnarray*}
N_{1m}&=&\int_{0}^{1}...\int_{0}^{1}a^{2}R_{m}^{2}(E_{m}^{2}+O_{m}^{2})dr_{1}...dr_{n-1}\\
N_{2m}&=&\int_{0}^{1}...\int_{0}^{1}S_{m}^{2}(r_{l-1}^{2}(E_{m}^{2}+O_{m}^{2})+a^{2}-r_{l-1}^{2})dr_{1}...dr_{n-1}\hspace{0.5cm}\text{(for any)}l=2,3,....,n.
\end{eqnarray*}
Further $N_{1m}=(n-1)N_{2m}-\frac{1}{4}(n-2)$. Now choose
\begin{equation}
R_{m}=\frac{1}{\sqrt{N_{1m}}}\hspace{.2cm}\text{and}\hspace{.2cm}S_{m}=\frac{1}{\sqrt{N_{2m}}}.\label{un10}
\end{equation}
Then we obtain
$$\sum_{q=1}^{n}\int_{D\times...\times D}\mid\mathcal Z,q\rangle\langle\mathcal Z,q\mid d\mu=\mathbb I_{n}\otimes\sum_{m=1}^{\infty}\mid\psi_{m} \rangle\langle\psi_{m}\mid .$$
Now let
\begin{equation}
T=\mathbb I_{n}\otimes\sum_{m=1}^{\infty}\mid\psi_{m} \rangle\langle\psi_{m}\mid ,\label{un11}
\end{equation}
which is a bounded invertible operator with $Ker T=\{0\}$, for
$$KerT=\{\Phi\in\tilde{\mathfrak H}\mid T\Phi =0\}.$$
Every vector $\Phi$ in $\tilde{\mathfrak H}$ can be written as
$$\Phi=\left(\sum_{k=1}^{\infty}\alpha_{k1}\phi_{k},\sum_{k=1}^{\infty}\alpha_{k2}\phi_{k},....\sum_{k=1}^{\infty}\alpha_{kn}\phi_{k}\right)^{T}$$
Thus $T\Phi=0$ gives
$$\text{diag}\left(\sum_{m=1}^{\infty}\frac{1}{(m+1)^{2}}\mid\phi_{m} \rangle\langle\phi_{m}\mid,\sum_{m=1}^{\infty}\frac{1}{(m+1)^{2}}\mid\phi_{m} \rangle\langle\phi_{m}\mid
,....\sum_{m=1}^{\infty}\frac{1}{(m+1)^{2}}\mid\phi_{m} \rangle\langle\phi_{m}\mid\right)\cdot$$
$$\hspace{2cm}
\left(\sum_{k=1}^{\infty}\alpha_{k1}\phi_{k},\sum_{k=1}^{\infty}\alpha_{k2}\phi_{k},....\sum_{k=1}^{\infty}\alpha_{kn}\phi_{k}\right)^{T}
=(0,0,....0)$$
Hence, for each $i$, we get
$$\sum_{m=1}^{\infty}\frac{1}{(m+1)^{2}}\alpha_{mi}\phi_{m}=0.$$
With the fact that $\{\phi_{m}\}_{m=0}^{\infty}$ is an orthonormal basis, for each $m$ and for each $i$, we have $\alpha_{mi}=0$. Thus we have $KerT=\{0\}$. The inverse of $T$ exists and the domain of $T^{-1}$, $\mathcal D({T^{-1}})$ is dense in $\tilde{\mathfrak H}$. For $\phi\in\mathcal D(T^{-1})$, consider
$$T(T^{-1}\phi)=\left[\sum_{q=1}^{n}\int_{D}\mid\mathcal Z ,q\rangle\langle\mathcal Z,q\mid d\mu\right]T^{-1}\phi.$$
That is
$$\phi=\sum_{q=1}^{n}\int_{D}\mid\mathcal Z,q \rangle\langle\mathcal Z,q\mid T^{-1}\phi\rangle d\mu.$$
Thus, for the vectors in $\mathcal D(T^{-1})$ we have a proper decomposition in the above sense. Now let $\phi\in\tilde{\mathfrak H}$, then there exists a sequence $\{\eta_{m}\}_{m=0}^{\infty}\subset\mathcal D(T^{-1})$ such that $\eta_{m}\rightarrow\phi$ as $m\rightarrow\infty$ in $\tilde{\mathfrak H}$. Further for each $\eta_{m}$ we have,
$$\eta_{m}=\sum_{q=1}^{n}\int_{D}\mid\mathcal Z ,q\rangle\langle\mathcal Z,q\mid T^{-1}\eta_{m}\rangle d\mu.$$
Now by taking limit both sides as $m\rightarrow\infty$, we can have
$$\phi=\lim_{m\rightarrow\infty}\left(\sum_{q=1}^{n}\int_{D}\mid\mathcal Z,q \rangle\langle\mathcal Z,q\mid T^{-1}\eta_{m}\rangle d\mu\right).$$
In this regard, we have a week decomposition for the vectors in $\tilde{\mathfrak H}-\mathcal D(T^{-1})$. Even though we do not have a complete resolution of the identity the CS form a frame with the frame operator $T$ (see \cite{key2}). The inverse operator, $T^{-1}$ is unbounded.
\section{Example: Vector coherent states with SU(1,1)}
In this section we build vector coherent states on $ D$ by considering it as the homogeneous space of the group $SU(1,1)$. In order to introduce the concept we need the following preliminaries.\\
The non-compact group $SU(1,1)$ is defined as,
$$SU(1,1)=\left\{g|g=\left( \begin{array}{cc}
\alpha&\beta\\
\overline{\beta}&\overline{\alpha}
\end{array}\right);\alpha , \beta\in\mathbb C , \det g=|\alpha|^{2}-|\beta|^{2}=1\right\}$$
and its maximal compact subgroup $K$ is given by
$$K=\left\{k|k=\left(\begin{array}{cc}e^{\frac{i\phi}{2}}&0\\0&e^{\frac{-i\phi}{2}}\end{array}\right);0\leq\phi\leq 2\pi\right\}.$$
The Cartan decomposition of an arbitrary $g\in SU(1,1)$ is,
$$g=\left( \begin{array}{cc}
\alpha&\beta\\
\overline{\beta}&\overline{\alpha}
\end{array}\right)=|\alpha|
\left(\begin{array}{cc}1&z\\ \overline{z}&1\end{array}\right)
\left(\begin{array}{cc}\frac{\alpha}{|\alpha|}&0\\0&\frac{\overline{\alpha}}{|\alpha|}\end{array}\right)$$
where $z=\beta\alpha^{-1}$ and $|\alpha|=(1-|z|^{2})^{-\frac{1}{2}}$.\\
Thus the coset space $SU(1,1)/K$ can be identified with the unit disc. Further it is known that the measure
$$d\nu(z,\overline{z})=\frac{1}{2\pi i}\frac{dz\wedge d\overline{z}}{(1-|z|^{2})^{2}}$$
on $D$ is invariant under the action of $SU(1,1)$. In polar coordinates
\begin{eqnarray*}
\mathcal Z=\left(\begin{array}{cc}1&z\\ \overline{z}&1\end{array}\right)=\left(\begin{array}{cc}1&re^{i\theta}\\ re^{-i\theta}&1\end{array}\right)\\
d\nu(z,\overline{z})=d\nu(r,\theta)=\frac{rdrd\theta}{\pi(1-r^{2})^{2}}
\end{eqnarray*}
where $0\leq r<1$ and $0\leq\theta\leq 2\pi$. Let $\{\chi^{1},\chi^{2}\}$
be the natural basis of $\mathbb C^{2}$. Now, as before, form the Hilbert space $\widetilde{\mathfrak H}=\mathbb C^{2}\otimes\mathfrak H$ and take $\{\Phi_{qm}=\chi^{q}\otimes \psi_{m}\mid q=1,2; m=0,1,...,\infty\}$ as a basis of it. With the above set up we form the set of coherent states in $L^{2}(D,d\mu)$ as,
$$|\mathcal Z,q\rangle=\mathcal N(|\mathcal Z|)\sum_{m=0}^{\infty}R_{m}\mathcal Z^{m}\Phi_{qm},\hspace{1cm}q=1,2$$
by suitably choosing the number $\mathcal N=\mathcal N(|\mathcal Z|)$ and a $2\times2$ matrix $R(m)$.\\
As we did before through diagonalization we obtain
$$\mathcal Z^{m}=\left(\begin{array}{cc}
E_{m}&O_{m}e^{i\theta}\\
O_{m}e^{i\theta}&E_{m}
\end{array}\right)$$
with
\begin{eqnarray*}
E_{m}&=&\frac{1}{2}[(1+r)^{m}+(1-r)^{m}]\\
O_{m}&=&\frac{1}{2}[(1+r)^{m}-(1-r)^{m}].
\end{eqnarray*}
Take
$$R(m)=\left(\begin{array}{cc}
R_{m}e^{im\theta}&0\\
0&R_{m}e^{im\theta}
\end{array}\right).$$
Now with the above choices, matrix multiplication yields
\begin{eqnarray}
\mid\mathcal Z,1\rangle&=&\mathcal N\sum_{m=0}^{\infty}
\left(\begin{array}{c}
R_{m}e^{im\theta}E_{m}\psi_{m}\\
R_{m}e^{im\theta}O_{m}\psi_{m}
\end{array}\right)\label{un12}\\
\mid\mathcal Z,2\rangle&=&\mathcal N\sum_{m=0}^{\infty}
\left(\begin{array}{c}
R_{m}e^{im\theta}O_{m}\psi_{m}\\
R_{m}e^{im\theta}E_{m}\psi_{m}
\end{array}\right)\label{un13}
\end{eqnarray}
First let us calculate the following finite integrals
\begin{eqnarray*}
\int_{0}^{1}E_{m}^{2}rdr&=&\int_{0}^{1}\frac{1}{4}(1+r)^{2m}+\frac{1}{2}(1-r^{2})^{m}+\frac{1}{4}(1-r)^{2m}rdr\\
&=&\frac{1}{4}\left(\sum_{j=0}^{2m}\left(\begin{array}{c}2m\\j\end{array}\right)\frac{1}{j+2}+\sum_{j=0}^{2m}\left(\begin{array}{c}2m\\j\end{array}\right)\frac{(-1)^{j}}{j+2}\right)+\frac{1}{4(m+1)}\\
&=&\frac{1}{2}\frac{4^{m}+1+m}{(2m+1)(m+1)}
\end{eqnarray*}
and
\begin{eqnarray*}
\int_{0}^{1}O_{m}^{2}rdr&=&\int_{0}^{1}\frac{1}{4}(1+r)^{2m}-\frac{1}{2}(1-r^{2})^{m}+\frac{1}{4}(1-r)^{2m}rdr\\
\end{eqnarray*}
\begin{eqnarray*}
&=&\frac{1}{4}\left(\sum_{j=0}^{2m}\left(\begin{array}{c}2m\\j\end{array}\right)\frac{1}{j+2}+\sum_{j=0}^{2m}\left(\begin{array}{c}2m\\j\end{array}\right)\frac{(-1)^{j}}{j+2}\right)-\frac{1}{4(m+1)}\\
&=&\frac{m}{2}\frac{4^{m}-1}{(2m+1)(m+1)}.
\end{eqnarray*}
With the above two integrals we can also have
$$\int_{0}^{1}(E_{m}^{2}+O_{m}^{2})rdr=\frac{2\times 4^{m}m+1}{2(2m+1)(m+1)}.$$
Further notice that $\langle\psi_{m}\mid\psi_{m}\rangle=\frac{1}{(1+m)^{2}}$.
Simple calculations yields
\begin{eqnarray}
\langle\mathcal Z,1\mid\mathcal Z,1\rangle&=&\sum_{m=0}^{\infty}\frac{1}{(1+m)^{2}}\int_{0}^{2\pi}\int_{0}^{1}\mathcal N^{2}R_{m}^{2}(E_{m}^{2}+O_{m}^{2})\frac{rd\theta dr}{\pi (1+r^{2})^{2}}\label{un14}\\
\langle\mathcal Z,2\mid\mathcal Z,2\rangle&=&\sum_{m=0}^{\infty}\frac{1}{(1+m)^{2}}\int_{0}^{2\pi}\int_{0}^{1}\mathcal N^{2}R_{m}^{2}(E_{m}^{2}+O_{m}^{2})\frac{rd\theta dr}{\pi (1+r^{2})^{2}}.\label{un15}
\end{eqnarray}
Now let us choose
\begin{eqnarray}
\mathcal N&=&\frac{\sqrt{6}}{\pi}(1+r^{2})\label{un16}\\
R_{m}&=&\sqrt{\frac{(2m+1)(m+1)}{2\times 4^{m}m+1}}.\label{un17}
\end{eqnarray}
With these choices we have
\begin{eqnarray*}
\langle\mathcal Z,1\mid\mathcal Z,1\rangle=\langle\mathcal Z,2\mid\mathcal Z,2\rangle&=&\sum_{m=0}^{\infty}\int_{0}^{2\pi}\int_{0}^{1}\frac{6}{\pi^{2}}\frac{(2m+1)(m+1)}{2\times 4^{m}m+1}(E_{m}^{2}+O_{m}^{2})rd\theta dr\\
&=&\sum_{m=0}^{\infty}\frac{1}{(1+m)^{2}}\frac{6}{\pi^{2}}\\
&=&1
\end{eqnarray*}
Let us turn our attention to the resolution of identity with a calculation of
\begin{eqnarray*}
\mid\mathcal Z,1\rangle\langle\mathcal Z,1\mid&=&\sum_{m=0}^{\infty}\sum_{k=0}^{\infty}\mathcal N^{2}\mid R_{m}\mathcal Z^{m}\Phi_{1m}\rangle\langle R_{k}\mathcal Z^{k}\Phi_{1k}\mid\\
&=&\sum_{m=0}^{\infty}\sum_{k=0}^{\infty}\mathcal N^{2}R_{m}\mathcal Z^{m}\mid\Phi_{1m}\rangle\langle\Phi_{1k}\mid\overline{ R_{k}\mathcal Z^{k}}^{T}\\
&=&\sum_{m=0}^{\infty}\sum_{k=0}^{\infty}\mathcal N^{2}
\left(\begin{array}{cc}R_{m}e^{im\theta}&0\\0&R_{m}e^{im\theta}\end{array}\right)
\left(\begin{array}{cc}E_{m}&O_{m}e^{i\theta}\\O_{m}e^{-i\theta}&E_{m}\end{array}\right)\times\\
&&\hspace{1cm}\left(\begin{array}{cc}1&0\\0&0\end{array}\right)
\left(\begin{array}{cc}E_{m}&O_{m}e^{i\theta}\\O_{m}e^{-i\theta}&E_{m}\end{array}\right)
\left(\begin{array}{cc}R_{m}e^{-im\theta}&0\\0&R_{m}e^{-im\theta}\end{array}\right)\\
&=&\sum_{m=0}^{\infty}\sum_{k=0}^{\infty}\mathcal N^{2}
\left(\begin{array}{cc}R_{m}R_{k}E_{m}E_{k}e^{i\theta(m-k)}&R_{m}R_{k}E_{m}O_{k}e^{i\theta(m+k+1)}
\\R_{m}R_{k}O_{m}E_{k}e^{-i\theta(m+k+1)}&R_{m}R_{k}O_{m}O_{k}e^{i\theta(m-k)}\end{array}\right)
\end{eqnarray*}
Thus we get
$$\int_{D}W\mid\mathcal Z,1\rangle\langle\mathcal Z,1\mid d\mu=\left(\begin{array}{cc}A&0\\0&B\end{array}\right)$$
where
\begin{eqnarray}
A&=&\sum_{m=0}^{\infty}\int_{0}^{2\pi}\int_{0}^{1}\mathcal N^{2}R_{m}^{2}E_{m}^{2}W\frac{rd\theta dr}{\pi (1+r^{2})^{2}}\otimes \mid\psi_{m}\rangle\langle\psi_{m}\mid\label{un18}\\
B&=&\sum_{m=0}^{\infty}\int_{0}^{2\pi}\int_{0}^{1}\mathcal N^{2}R_{m}^{2}O_{m}^{2}W\frac{rd\theta dr}{\pi (1+r^{2})^{2}}\otimes\mid\psi_{m}\rangle\langle\psi_{m}\mid.\label{un19}
\end{eqnarray}
Similarly we get
$$\int_{D}W\mid\mathcal Z,2\rangle\langle\mathcal Z,2\mid d\mu=\left(\begin{array}{cc}B&0\\0&A\end{array}\right).$$
With all these we have
\begin{equation}
\int_{D}W\mid\mathcal Z,1\rangle\langle\mathcal Z,1\mid d\mu+\int_{D}W\mid\mathcal Z,2\rangle\langle\mathcal Z,2\mid d\mu=\left(\begin{array}{cc}A+B&0\\0&A+B\end{array}\right).\label{un20}
\end{equation}
Now let us calculate $A+B$
$$A+B=\sum_{m=0}^{\infty}\int_{0}^{2\pi}\int_{0}^{1}\mathcal N^{2}R_{m}^{2}(E_{m}^{2}+O_{m}^{2})W\frac{rd\theta dr}{\pi (1+r^{2})^{2}}\otimes\mid\psi_{m}\rangle\langle\psi_{m}\mid$$
In order to get
$$\int_{0}^{2\pi}\int_{0}^{1}\mathcal N^{2}R_{m}^{2}(E_{m}^{2}+O_{m}^{2})W\frac{rd\theta dr}{\pi (1+r^{2})^{2}}=1$$
let us choose
$$W=\frac{\pi^{2}}{6}.$$
With this choice we have
\begin{eqnarray*}
&&\int_{0}^{2\pi}\int_{0}^{1}\mathcal N^{2}R_{m}^{2}(E_{m}^{2}+O_{m}^{2})W\frac{rd\theta dr}{\pi (1+r^{2})^{2}}\\
&=&\int_{0}^{2\pi}\int_{0}^{1}\frac{6(1+r^{2})^{2}}{\pi^{2}}\frac{(2m+1)(m+1)}{2\times 4^{m}m+1}(E_{m}^{2}+O_{m}^{2})\frac{\pi^{2}}{6}\frac{rd\theta dr}{\pi (1+r^{2})^{2}}\\
&=&\frac{2\times 4^{m}m+1}{2(2m+1)(m+1)}\frac{6}{\pi^{2}}\frac{(2m+1)(m+1)}{2\times 4^{m}m+1}\frac{\pi^{2}}{6}2\pi\\
&=&1
\end{eqnarray*}
Now we have
\begin{equation}
A+B=\sum_{m=0}^{\infty}\mid\psi_{m}\rangle\langle\psi_{m}\mid\label{un21}
\end{equation}
which yields
$$\int_{D}W\mid\mathcal Z,1\rangle\langle\mathcal Z,1\mid d\mu+\int_{D}W\mid\mathcal Z,2\rangle\langle\mathcal Z,2\mid d\mu=\mathbb I_{2}\otimes\sum_{m=0}^{\infty}\mid\psi_{m}\rangle\langle\psi_{m}\mid .$$
Let 
$$T=\mathbb I_{2}\otimes\sum_{m=0}^{\infty}\mid\psi_{m}\rangle\langle\psi_{m}\mid .$$
Thus we have an operator $T$ as in equation (\ref{un11}) with $n=2$. The decomposition of any vector $\phi$ in $\tilde{\mathfrak H}$ follows by replacing $n=2$ in the discussion which we have right after equation (\ref{un11}).

\section{Remarks and Discussions}
In this section we will discuss some other possibilities of our choices over the construction and resulting difficulties.
\begin{enumerate}
\item[$\bullet$]
We have chosen a basis of the Hilbert space in an unusual way as $\psi_{m}=\frac{\phi_{m}}{m+1}$. We explain the convenience of this choice using the $SU(1,1)$ example. Instead of this choice if we take $\phi_{m}$, in equations (\ref{un12}) and (\ref{un13}) we will have $\phi_{m}$ instead of $\psi_{m}$, which will change equations (\ref{un14}) and (\ref{un15}) as
\begin{eqnarray}
\langle\mathcal Z,1\mid\mathcal Z,1\rangle&=&\sum_{m=0}^{\infty}\int_{0}^{2\pi}\int_{0}^{1}\mathcal N^{2}R_{m}^{2}(E_{m}^{2}+O_{m}^{2})\frac{rd\theta dr}{\pi (1+r^{2})^{2}}\label{un22}\\
\langle\mathcal Z,2\mid\mathcal Z,2\rangle&=&\sum_{m=0}^{\infty}\int_{0}^{2\pi}\int_{0}^{1}\mathcal N^{2}R_{m}^{2}(E_{m}^{2}+O_{m}^{2})\frac{rd\theta dr}{\pi (1+r^{2})^{2}}.\label{un23}
\end{eqnarray}
Further, in equation (\ref{un20}) $\psi_{m}$ will be replaced by $\phi_{m}$, i.e,
\begin{equation}
A+B=\sum_{m=0}^{\infty}\mid\phi_{m}\rangle\langle\phi_{m}\mid\int_{0}^{2\pi}\int_{0}^{1}\mathcal N^{2}R_{m}^{2}(E_{m}^{2}+O_{m}^{2})W\frac{rd\theta dr}{\pi (1+r^{2})^{2}}.\label{un24}
\end{equation}
So in (\ref{un24}) by choosing
$$
\mathcal N = \frac{(1+r^{2})}{\sqrt{2}},\hspace{1cm}
R_{m}=\frac{1}{\sqrt{N_{m}}},\hspace{1cm}\text{and}\hspace{1cm}
W=1
$$
with
$$N_{m}=\int_{0}^{1}r(E_{m}^{2}+O_{m}^{2})dr=\frac{2\times 4^{m}m+1}{2(2m+1)(m+1)}$$
we can have the resolution of identity. But these choices will make the series in (\ref{un22}) and (\ref{un23}) diverge. In order to have the convergence in (\ref{un22}) and (\ref{un23}), at least we have to have
$$R_{m}^{2}\cdot\frac{2\times 4^{m}m+1}{2(2m+1)(m+1)}\sim\frac{1}{m^{p}}$$
with $p>1$. If we make such a choice for $R_{m}$, our moment problem will be
$$\int_{0}^{1}(E_{m}^{2}+O_{m}^{2})Wrdr=\frac{2\times 4^{m}m+1}{2(2m+1)(m+1)}m^{p}.$$
We have experienced difficulty in solving this moment problem by taking $W$ as a function of $r$ only, but it can be solved if we take $W=f(r,m)$. Generally, as a weight function, dependence of $W$ on $m$ is not allowed. In this regard, we made the choice $\psi_{m}=\frac{\phi_{m}}{m+1}$ . Notice also that $\psi_{m}$ can be chosen in many ways in a similar fashion.
\item[$\bullet$]
We have obtained proper decomposition  for the vectors in our intended space up to a dense subset. The effort of getting the decomposition for the remaining nowhere dense set is limited by the unboundedness of the inverse operator $T^{-1}$. In this case, we have an approximation for the decomposition.

\item[$\bullet$]
The $SU(1,1)$ example presented here can be considered as a preliminary step of constructing vector coherent states on classical domains. The construction  directly depends on the explicit form of the elements of the classical groups. As we know, many of the matrix groups consist elements with an unpleasant explicit form in terms of calculations. In this regard, construction of vector coherent states on other classical domains may need little more effort.
\end{enumerate}

                      \end{document}